\documentclass[10pt,twocolumn,twoside]{IEEEtran}
\usepackage{amsfonts, amssymb, amsmath, amsthm, bm, bbm, dsfont}
\usepackage{url, color, upgreek, cite, epsfig, psfrag, rotating}
\usepackage{algpseudocode}

\newcommand{\Rbb}{\mathbb{R}} \newcommand{\Cbb}{\mathbb{C}}
\newcommand{\E}{{\mathbb{E}}} 
\newcommand{\scp}[2]{\langle #1, #2 \rangle}
\newtheorem{theorem}{Theorem} \newtheorem{definition}{Definition}
\newtheorem{proposition}{Proposition} \newtheorem{lemma}{Lemma}
 
\newcommand{\inv}[1]{\frac{1}{#1}} 
 \newcommand{\sign}{{\rm sgn}\,}
\renewcommand{\leq}{\leqslant} \renewcommand{\geq}{\geqslant}
\newcommand{\norm}[1]{\|#1\|} \newcommand{\abs}[1]{\left| #1 \right|}
\newcommand{\ma}[1]{\mathsf{#1}}\newcommand{\set}[1]{\mathcal{#1}}
\DeclareMathOperator*{\argmin}{arg\,min}

\title{On Variable Density Compressive Sampling}

\author{Gilles~Puy, Pierre~Vandergheynst, and Yves~Wiaux 
\thanks{Copyright (c) 2011 IEEE. Personal use of this material is permitted. However, permission to use this material for any other purposes must be obtained from the IEEE by sending a request to pubs-permissions@ieee.org.}
\thanks{G.~Puy, P.~Vandergheynst, and Y.~Wiaux are with the Institute of Electrical Engineering,  Ecole Polytechnique F{\'e}d{\'e}rale de Lausanne (EPFL), CH-1015 Lausanne, Switzerland. G.~Puy is also with the Institute of the Physics of Biological Systems, Ecole Polytechnique F{\'e}d{\'e}rale de Lausanne (EPFL), CH-1015 Lausanne, Switzerland. Y.~Wiaux is also with the Institute of Bioengineering,  Ecole Polytechnique F{\'e}d{\'e}rale de Lausanne (EPFL), CH-1015 Lausanne, Switzerland, and with the Department of Radiology and Medical Informatics, University of Geneva (UniGE), CH-1211 Geneva, Switzerland. E-mail: gilles.puy@epfl.ch; pierre.vandergheynst@epfl.ch; yves.wiaux@epfl.ch - Address: EPFL STI IEL LTS2 - ELE 227 - Station 11 - CH-1015 Lausanne} 
\thanks{This work is supported in part by the Center for Biomedical Imaging of the Geneva and Lausanne Universities, EPFL, and the Leenaards and Louis-Jeantet foundations, also by the Swiss National Science Foundation under grant PP00P2-123438, and by the EPFL-Merck Serono Alliance award.}}

\begin{document}
\maketitle

\begin{abstract}
Incoherence between sparsity basis and sensing basis is an essential concept for compressive sampling. In this context, we advocate a coherence-driven optimization procedure for variable density sampling. The associated minimization problem is solved by use of convex optimization algorithms. We also propose a refinement of our technique when prior information is available on the signal support in the sparsity basis. The effectiveness of the method is confirmed by numerical experiments. Our results also provide a theoretical underpinning to state-of-the-art variable density Fourier sampling procedures used in MRI.
\end{abstract}

\begin{IEEEkeywords}
compressed sensing, variable density sampling, magnetic resonance imaging.
\end{IEEEkeywords}

%
%
%
%
\section{Introduction}
\label{sec:introduction}

Compressed sensing demonstrates that sparse signals can be sampled through linear and non-adaptive measurements at a sub-Nyquist rate, and still accurately recovered by means of non-linear iterative algorithms. The theory requires incoherence between the sensing and sparsity bases and a lot of work has thus been dedicated to design such sensing systems \cite{candes07}.

In the present work, we concentrate on $s$-sparse digital signals $\bm{\alpha} = \left(\alpha_i\right)_{1\leq i \leq N} \in \Cbb^N$ in an orthonormal basis $\ma \Psi = (\bm \psi_1, ..., \bm \psi_N) \in \Cbb^{N \times N}$. The vector $\bm{\alpha}$ contains $s$ non-zero entries and its support is defined as $S = \left\{i : \abs{\alpha_i} > 0, 1 \leq i \leq N \right\}$. We denote $\bm{\alpha}_S \in \Cbb^s$ the vector made of the $s$ non-zero entries of $\bm{\alpha}$. This signal is probed by projection onto $m$ vectors of another orthonormal basis $\ma{\Phi} = (\bm \phi_1, ..., \bm \phi_N) \in \Cbb^{N \times N}$. The indices of the selected vectors are denoted $\Omega = \left\{l_1, \ldots, l_m\right\}$ and $\ma \Phi^\dagger_\Omega$ is the $m \times N$ matrix made of the selected rows of $\ma \Phi^\dagger$, where the symbol $\cdot^\dagger$ stands for the conjugate transpose operation. The measurement vector $\bm y \in \Cbb^m$ thus reads as
\begin{eqnarray}
\label{eq:measurement model}
\bm y = \ma{A}_\Omega \, \bm \alpha \textnormal{ with } \ma{A}_\Omega = \ma{\Phi}^\dagger_\Omega \ma{\Psi} \in \Cbb^{m \times N}.
\end{eqnarray}
We also denote $\ma{A} = \ma{\Phi}^\dagger\ma{\Psi} \in \Cbb^{N \times N}$. Finally, we aim at recovering $\bm{\alpha}$ by solving the $\ell_1$-minimization\footnote{$\norm{\bm{\alpha}}_1 = \sum_{1 \leq i\leq N} \abs{\alpha_i}$ ($\abs{\cdot}$ denotes the complex magnitude).} problem
\begin{eqnarray}
\label{eq:BP}
\hat{\bm{\alpha}} = \argmin_{\bm{\alpha} \in \Cbb^N} \norm{\bm{\alpha}}_1 \text{  subject to  }  \bm y=\ma{A}_\Omega \bm{\alpha}.
\end{eqnarray}

In this setting, common strategies focus on uniform random selection of the indices $l_1, \ldots, l_m$. For signals sparse in the Dirac basis, a uniform random selection of Fourier basis vectors represents the best sampling strategy. Indeed, the Dirac and Fourier basis are optimally incoherent. Natural signals are however rather sparse in multi-scale bases, e.g. wavelet bases, not optimally incoherent with the Fourier basis. Many measurements are thus needed to reconstruct such signals accurately. This is for example the case in magnetic resonance imaging (MRI). To reduce the number of measurements, the authors in \cite{lustig07} rely on the fact that the energy of MRI signals is essentially concentrated at low frequencies. They thus propose to select Fourier basis vectors according to a variable density sampling profile selecting more low frequencies than high frequencies. This approach was shown to drastically enhance the quality of the reconstructed signals. This method is however essentially empirical and the reconstruction quality depends on the shape of the sampling profile used. Let us also mention that a line of justification for VDS was proposed in terms of the variable sparsity of the signals of interest as a function of scale in a wavelet sparsity basis \cite{candes07, wang10}.

In this letter, we study VDS in the theoretical framework of compressed sensing. In Section \ref{sec:variable density sampling}, we describe the latest compressed sensing results for sparse signals probed in bounded orthonormal system, and explain how they encompass variable density sampling procedures. In Section \ref{sec:vds optimization}, we introduce a minimization problem for the coherence between the sparsity and sensing bases, whose solution provides an optimized sampling profile. This minimization problem is solved with the use of convex optimization algorithms. We also propose a further refinement of our technique when prior information is available on the signal support $S$. In Section \ref{sec:experiments}, we illustrate the effectiveness of the method through numerical simulations. We also provide a comparison of the Fourier VDS profile in the presence of prior information and corresponding reconstruction qualities, with the state-of-the-art VDS approaches used in MRI. Finally, we conclude in Section \ref{sec:conclusion}.

%
%
%
%
\section{Variable density sampling}
\label{sec:variable density sampling}

In the setting presented in Section \ref{sec:introduction}, the compressed sensing theory demonstrates that if the sampling indices $l_1, \ldots, l_m$ are chosen randomly and independently according to a discrete probability measure $P$ defined on $\left\{1, \ldots, N\right\}$, then a small number $m\ll N$ of random measurements are sufficient for an exact reconstruction of $\bm{\alpha}$ \cite{rauhut10}.
\begin{theorem}[Theorem $4.2$, \cite{rauhut10}]
\label{th:standard non-uniform recovery}
Let $\ma{A} = \ma{\Phi}^\dagger\ma{\Psi} \in \Cbb^{N \times N}$, and $\bm{\alpha} \in \Cbb^N$ be a $s$-sparse vector such that\footnote{$\sign\left(\bm{\alpha}_S\right) \in \Cbb^s$ is the $s$-dimensional vector with entries $\alpha_i/\abs{\alpha_i}$, $\forall i \in S$.} $\sign\left(\bm{\alpha}_S\right) \in \Cbb^s$ is a random Steinhaus sequence. Assume that the sampling indices $\Omega = \left\{l_1, \ldots, l_m\right\}$ are selected randomly and independently according to a discrete probability measure $P$ defined on $\left\{1, \ldots, N\right\}$. Let $\bm{y} = \ma{A}_\Omega \bm{\alpha} \in \Cbb^{m}$ and define 
\begin{eqnarray}
\label{eq:coherence}
\mu(P) = \inv{N^{1/2}} \, \max_{1\leq i,j \leq N} \frac{\left| \scp{\bm \phi_i}{\bm \psi_j} \right|}{P^{1/2}(i)}.
\end{eqnarray}
For a universal constant $C>0$, if
\begin{eqnarray}
\label{eq:recovery condition}
m \geq C N \mu^2(P) s \log^2(6N/\varepsilon),
\end{eqnarray}
then $\bm{\alpha}$ is the unique minimizer of the $\ell_1$-minimization problem (\ref{eq:BP}) with probability at least $1-\varepsilon$.
\end{theorem}
In the above theorem, the parameter $\mu(P)$ stands for the mutual coherence between the measurement basis $\ma{\Phi}$ and the sparsity basis $\ma{\Psi}$. This value depends on the probability measure $P$ and statisfies $\mu(P) \geq N^{-1/2}$ \cite{rauhut10}. The smaller the mutual coherence the smaller the required number of measurements for exact recovery.

Let us highlight that with the selection procedure described in Theorem \ref{th:standard non-uniform recovery}, the number of measurements is exactly $m$ but one measurement vector might be selected more than once. This characteristic is not always suitable in practical applications, such as MRI, particularly in a VDS configuration. Indeed, a sensing basis vector $\bm{\phi}_i$, whose associated probability of selection $P(i)$ is high, will be selected multiple times thus reducing the quantity of information probed. To avoid this phenomenon, we propose another selection process.

In the remainder, the sampling indices are selected according to an admissible sampling profile for $m$ measurements.
\begin{definition}[Admissible sampling profile] A vector $\bm{p} = \left(p_j\right)_{1\leq j \leq N} \in \Rbb^N$ is an admissible sampling profile for a number $m$ of measurements if $p_j \in (0, 1]$ for all $1 \leq j \leq N$, and $\norm{\bm{p}}_1 = m$. The set of all admissible sampling profiles for a number $m$ of measurements is denoted $\set{P}(m)$.
\end{definition}

Let $\bm{p} \in \set{P}(m)$ be an admissible sampling profile, the sampling indices are selected by generating a sequence $\left(\delta_1, ..., \delta_N\right) \in \Rbb^N$ of independent Bernouilli random variables taking value $0$ or $1$ and such that $\delta_i$ is equal to $1$ with probability $p_i$ for all $1\leq i \leq N$. The set of selected indices is then defined as $\Omega = \left\{l : \delta_l = 1\right\}$. With the proposed sampling strategy, one measurement vector can be selected only once. The constraint that $\norm{\bm{p}}_1 = m$ imposes that the number of measurements is $m$ on average over realizations of a sequence $\left(\delta_1, ..., \delta_N\right)$. Note that for $N \gg 1$, the variability of the number of measurements is negligible.

As suggested in \cite{rauhut10}, one can actually show that the recovery condition (\ref{eq:recovery condition}) still holds with the coherence
\begin{eqnarray}
\mu(\bm{p}) = \left(\frac{m}{N}\right)^{1/2} \max_{1\leq i, j \leq N} \frac{\left| \scp{\bm \phi_i}{\bm \psi_j} \right|}{p^{1/2}_i}.
\end{eqnarray}
The required elements of proof are provided in Appendix \ref{ap:proof coherence}.

%
%
%
%
\section{Sampling profile optimization}
\label{sec:vds optimization}

\begin{figure}
\centering
\psfrag{0}{\hspace{-0.5mm}\scriptsize$0$} \psfrag{0.1}{}
\psfrag{0.2}{\hspace{-1mm}\scriptsize$0.2$} \psfrag{0.3}{}
\psfrag{0.4}{\hspace{-1mm}\scriptsize$0.4$} \psfrag{0.5}{}
\psfrag{0.6}{\hspace{-1mm}\scriptsize$0.6$} \psfrag{0.7}{}
\psfrag{0.8}{\hspace{-1mm}\scriptsize$0.8$} \psfrag{0.9}{}
\psfrag{1}{\hspace{-1mm}\scriptsize$1$}
\psfrag{100}{} \psfrag{200}{\hspace{-1.5mm} \scriptsize$200$} \psfrag{448}{\hspace{-1.5mm} \scriptsize$448$}
\psfrag{300}{} \psfrag{400}{\hspace{-1.5mm} \scriptsize$400$} \psfrag{480}{\hspace{-1.5mm} \scriptsize$480$}
\psfrag{500}{} \psfrag{600}{\hspace{-1.5mm} \scriptsize$600$} \psfrag{512}{\hspace{-1.5mm} \scriptsize$512$}
\psfrag{700}{} \psfrag{800}{\hspace{-1.5mm} \scriptsize$800$} \psfrag{544}{\hspace{-1.5mm} \scriptsize$544$}
\psfrag{900}{} \psfrag{1000}{\hspace{-1.5mm} \scriptsize$1000$} \psfrag{576}{\hspace{-1.5mm} \scriptsize$576$}
\psfrag{m}{\hspace{-1.5mm} \scriptsize$m$} \psfrag{Prob}{\hspace{1.5mm}\scriptsize$\varepsilon$}
\psfrag{j}{\hspace{-1.5mm} \scriptsize $i$} \psfrag{pdf}{}
\includegraphics[width=4.1cm,keepaspectratio]{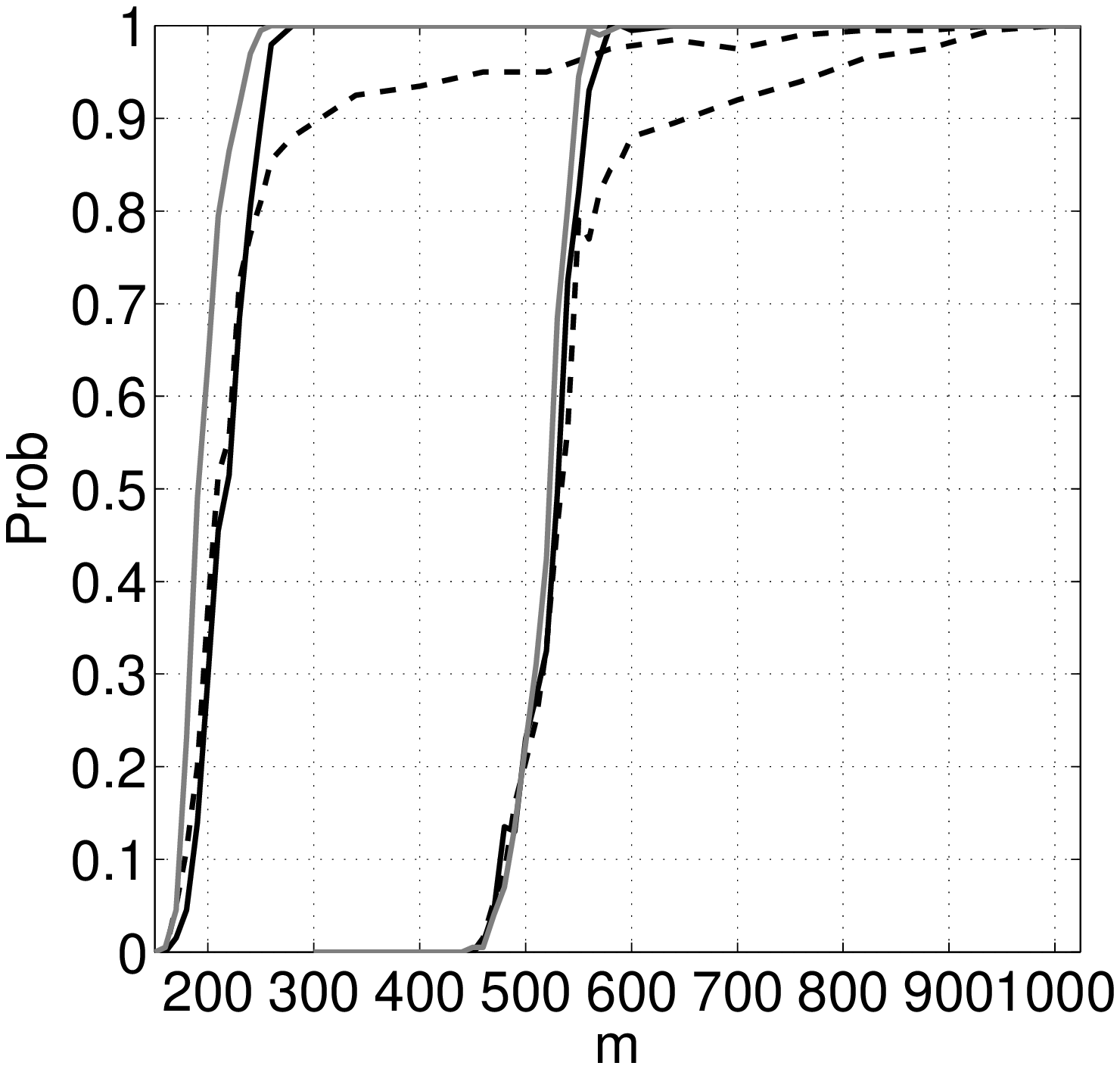}
\includegraphics[width=4.1cm,keepaspectratio]{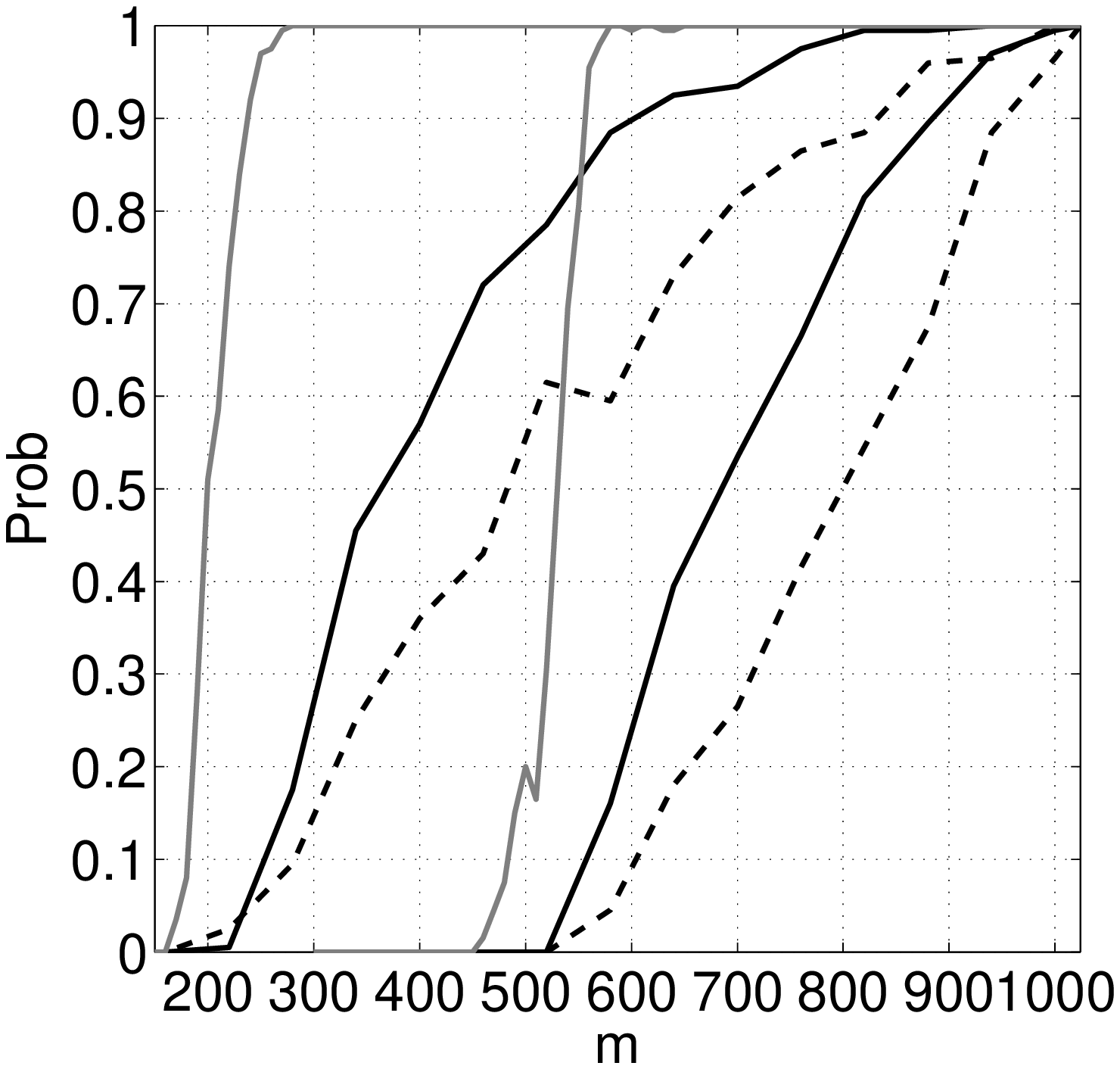}\\
\includegraphics[width=4.1cm,keepaspectratio]{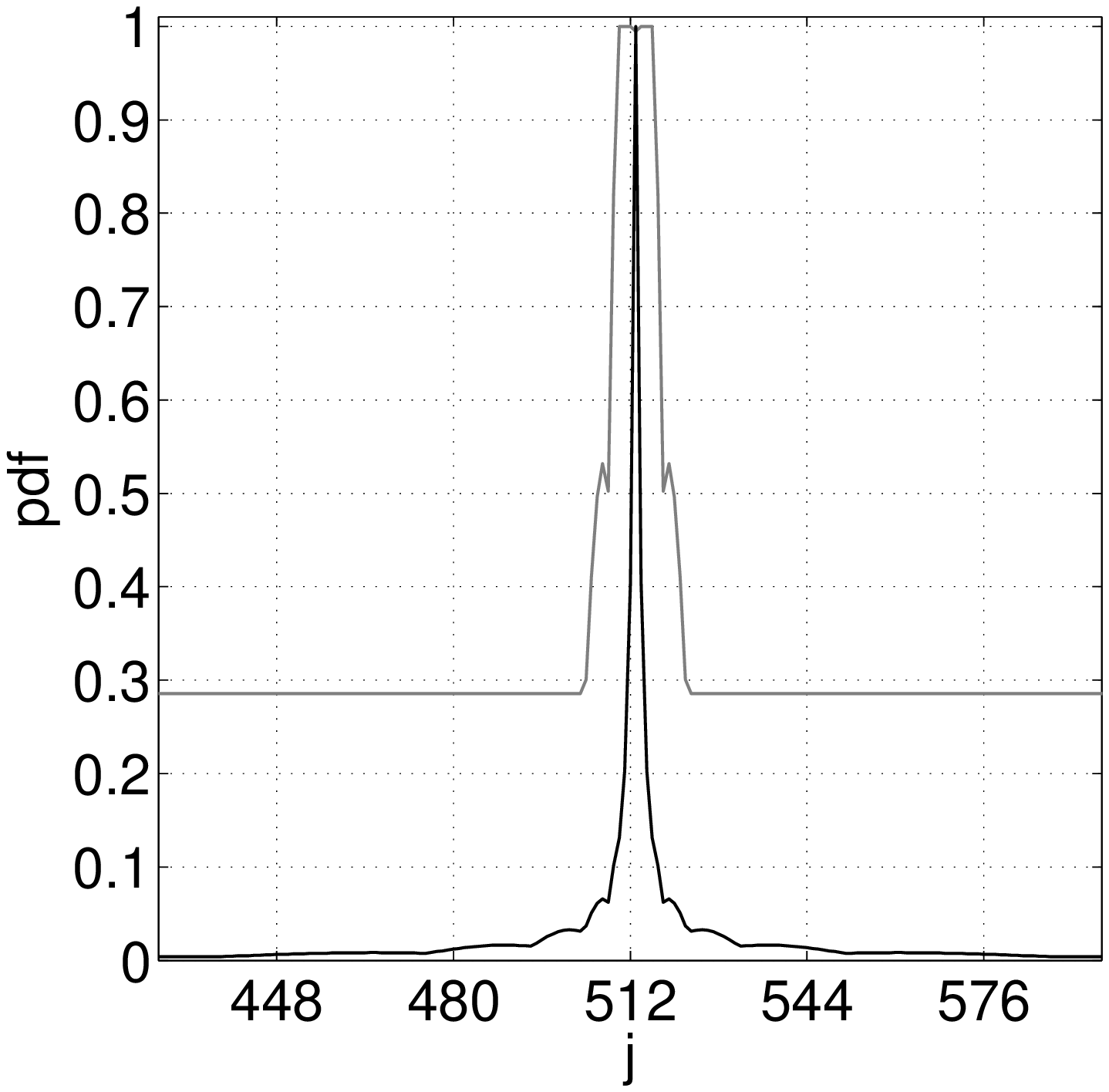}
\includegraphics[width=4.1cm,keepaspectratio]{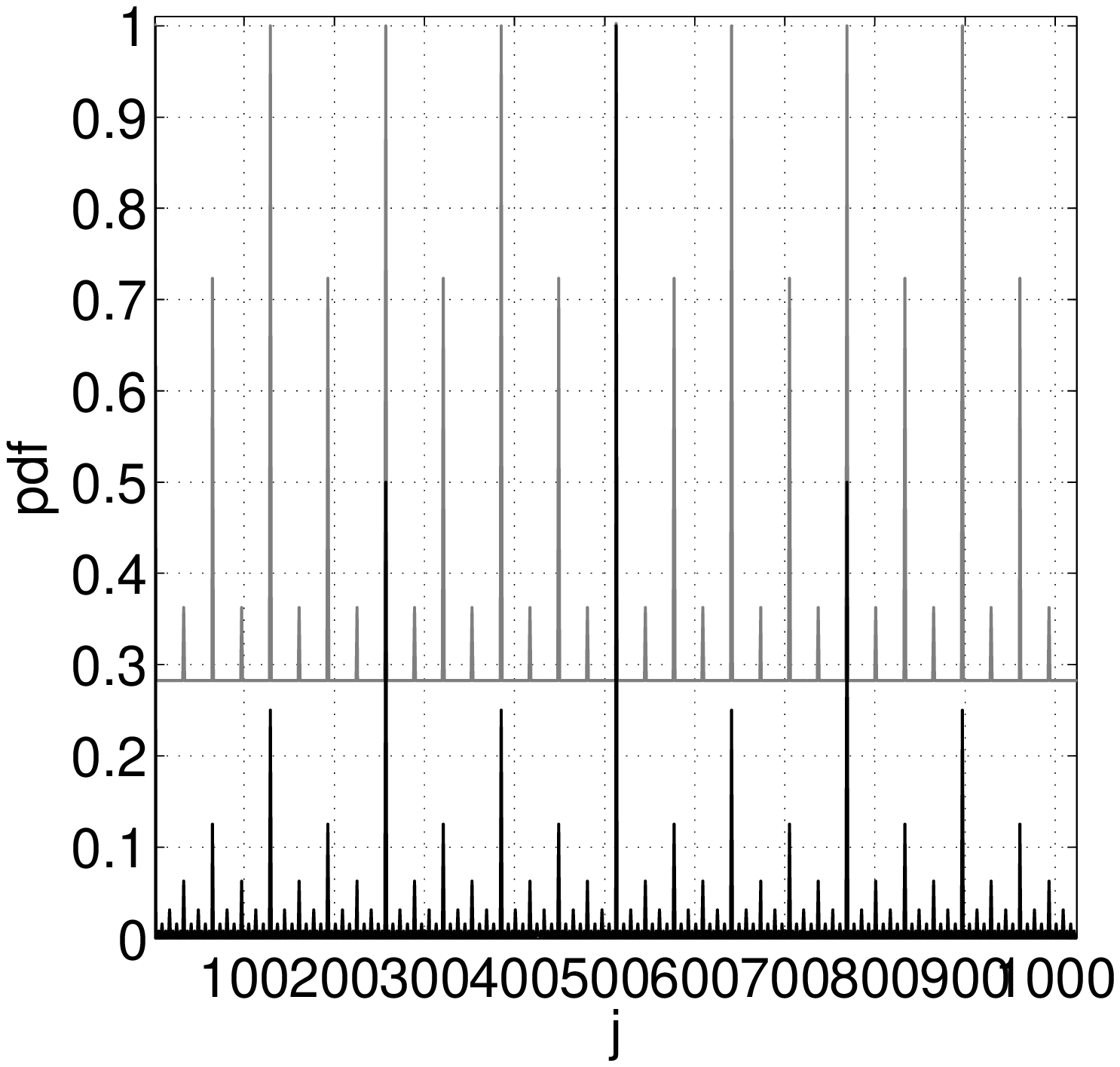}\\
\caption{\label{fig:probability recovery}Top panels: probability of recovery $\epsilon$ of $s$-sparse signals in the Haar wavelet basis ($N=1024$) as a function of the number of measurements $m$ in the Fourier basis (left panel) and in the Hadamard basis (right panel). The dark dashed, dark continuous, and light continuous curves show the probability of recovery with a uniform sampling, an optimized variable density sampling, and the spread spectrum technique respectively. Curves on the left correspond to $s=50$ and those on the right to $s=200$. Bottom panels: light curves show the optimized sampling profile for $m=300$ obtained with a sampling in the Fourier basis (left panel) and in the Hadamard basis (right panel). Dark curves show the values $\max_{1\leq j \leq N} \left| \scp{\bm \phi_i}{\bm \psi_j} \right|^2$ for all $1\leq i \leq N$.\vspace{-4mm}}
\end{figure} 

Let us assume that the number of measurements $m$ is fixed. In order to recover the highest sparsity $s$ possible, Theorem \ref{th:standard non-uniform recovery} shows that we should use the sampling profile $\bm{p} \in \set{P}(m)$ minimizing the mutual coherence $\mu(\bm{p})$. Therefore, we propose to solve the following optimization problem
\begin{eqnarray}
\label{eq:optimization problem}
(\hat{\bm{p}}, \hat{\bm{q}}) = \argmin_{(\bm{p}, \bm{q}) \in \Rbb^{N \times 2}} \norm{\ma{B}\,\bm{q}}_\infty + \lambda \norm{\bm{p} \cdot \bm{q} - \bm{1}}_2^2
\textnormal{ s.t. } \bm{p} \in \set{K}_{\tau},
\end{eqnarray}
where $\lambda \in [0, +\infty)$, $\tau \in (0, 1]$, $\set{K}_{\tau} = \{\bm{p} \in \left[\tau, 1\right]^N : \norm{\bm{p}}_1 \leq m\}$, $\bm{1} \in \Rbb^N$ is the vector with all its entries equals to $1$, $\bm{p} \cdot \bm{q}$ is the entry-by-entry multiplication between the vector $\bm{p}$ and $\bm{q}$, $\norm{\cdot}_2$ and $\norm{\cdot}_\infty$ are respectively the $\ell_2$-norm and $\ell_\infty$-norm\footnote{$\norm{\bm{x}}_2^2 = \sum_{1 \leq j \leq N} \abs{x_j}^2$ and $\norm{\bm{x}}_\infty = \max_{1 \leq j \leq N} \abs{x_j}$.}, and $\ma{B} \in \Cbb^{N \times N}$ is the diagonal matrix with entries $\max_{1\leq j \leq N} \left| \scp{\bm \phi_i}{\bm \psi_j} \right|^2$ on the diagonal, $1 \leq i \leq N$. 

In the above problem, the term $\norm{\bm{p} \cdot \bm{q} - \bm{1}}_2^2$ ensures that $p_i \simeq 1/q_i$ for all $1 \leq i \leq N$. The higher the value of the parameter $\lambda$ the further this constraint is enforced. In the limit where $\bm{p} \cdot \bm{q} = \bm{1}$, we have $\norm{\ma{B}\,\bm{q}}_\infty = \mu^2(\bm{p})$ confirming that problem (\ref{eq:optimization problem}) seeks to minimize the mutual coherence. Note that the minimization problem imposes that $\bm{p}$ belongs to the set $\set{K}_{\tau}$ which is different from the set $\set{P}(m)$. Consequently, we do not have necessarily $\norm{\bm{p}}_1 = m$. However, we note that in practice the constraint $\norm{\bm{p}}_1 \leq m$ is always saturated for high enough values of $\lambda$.

To solve problem (\ref{eq:optimization problem}), we adopt the following procedure:
\begin{algorithmic}[1]
\State Set $t=0$ and $\hat{\bm{p}}^{(0)} = (m/N)_{1 \leq j \leq m}$;
\Repeat
\State $\hat{\bm{q}}^{(t)} \gets \argmin_{\bm{q} \in \Rbb^{N}} \norm{\ma{B}\,\bm{q}}_\infty + \lambda \norm{\hat{\bm{p}}^{(t)} \cdot \bm{q} - \bm{1}}_2^2$;\label{eq:solve q}
\State $\hat{\bm{p}}^{(t+1)} \gets \argmin_{\bm{p} \in \Rbb^{N}} \norm{\bm{p} \cdot \hat{\bm{q}}^{(t)} - \bm{1}}_2^2 \textnormal{ s.t. } \bm{p} \in \set{K}_\tau$;\label{eq:solve p}
\State $t \gets t+1$;
\Until{convergence}
\end{algorithmic}

Subproblems at step \ref{eq:solve q}) and \ref{eq:solve p}) are convex problems. The subproblem at step \ref{eq:solve q}) is solved iteratively using a forward-backward algorithm and the one at step \ref{eq:solve p}) thanks to a parallel proximal algorithm \cite{bauschke11}. Both algorithms require the computation of simple proximity operators. The computation of the one corresponding to $\norm{\ma{B}\,\cdot}_\infty$ essentially reduces to a projection onto an $\ell_1$-ball (see Appendix \ref{ap:proof prox}). This projection, as well as the one onto the $\ell_1$-ball of radius $m$, can be computed using the method\footnote{Code available at \url{http://www.cs.ubc.ca/labs/scl/spgl1}} presented in \cite{berg08}. Note that for both subproblems, the computational complexity at each iteration is essentially driven by these projections for which the method in \cite{berg08} has a worst-case complexity of $\mathcal{O}(N \log N)$. For $N=1024$, as in the forthcoming experiments, the overall algorithm converges in at most a few seconds. Our procedure therefore easily scales to larger $N$.

When prior information is available on the signal support $S$, we can refine our technique to find a sampling profile adapted to this support. Indeed, if the signal support $S$ is known in advance then Theorem \ref{th:standard non-uniform recovery} applies with the coherence
\begin{eqnarray}
\mu(\bm{p}, S) = \left(\frac{m}{s N} \max_{1\leq i  \leq N} \frac{\sum_{j \in S} \left| \scp{\bm \phi_i}{\bm \psi_j} \right|^2}{p_i} \right)^{1/2}.
\end{eqnarray}
We let the reader refer to Appendix \ref{ap:proof coherence} and equation (\ref{eq:bound model}) for more details. An optimized sampling profile associated with the set $S$ can thus be obtained by substituting the diagonal matrix $\ma{C} \in \Rbb^{N \times N}$ with entries $s^{-1}\sum_{j \in S} \left| \scp{\bm \phi_i}{\bm \psi_j} \right|^2$ on the diagonal, $1 \leq i \leq N$, for the matrix $\ma{B}$ in problem (\ref{eq:optimization problem}).

%
%
%
%
\section{Experiments}
\label{sec:experiments}

In order to evaluate the proposed method in a general setting, we conduct two experiments. For the first one, we choose the Haar wavelet basis as the sparsity basis $\ma{\Psi}$ and the Fourier basis as the sensing basis $\ma{\Phi}$. We generate complex $s$-sparse signals of size $N=1024$ with $s \in \{50, 200\}$. The positions of the non-zero coefficients are chosen uniformly at random in $\left\{1, \ldots, N \right\}$, their phases are set by generating a Steinhaus sequence, and their amplitudes follows a uniform distribution over $\left[0, 1 \right]$. The signals are then probed according to relation (\ref{eq:measurement model}) and reconstructed from different number of measurements $m$ by solving the $\ell_1$-minimization problem (\ref{eq:BP}) with the SPGL$1$ toolbox \cite{berg08}. For each value of $m$, the selected sensing basis vectors are chosen using the method described in Section \ref{sec:variable density sampling} using either a uniform density profile or the profile $\hat{\bm{p}}$ obtained by solving problem (\ref{eq:optimization problem}) with $\lambda = 0.05$. Each time, the probability of recovery\footnote{Perfect recovery is considered if the $\ell_2$-norm between the original signal $\bm{x}$ and the reconstructed signal $\bm{x}^\star$ satisfies: $\norm{\bm{x}-\bm{x}^\star}_2\leq10^{-3}\norm{\bm{x}}_2$.} is computed over $200$ simulations. For the second experiment, the same setting is used but with the Hadamard basis as the sensing basis $\ma{\Phi}$.

\begin{figure}
\centering
\psfrag{0}{\hspace{-0.5mm}\scriptsize$0$} \psfrag{0.1}{}
\psfrag{0.2}{\hspace{-1mm}\scriptsize$0.2$} \psfrag{0.3}{}
\psfrag{0.4}{\hspace{-1mm}\scriptsize$0.4$} \psfrag{0.5}{}
\psfrag{0.6}{\hspace{-1mm}\scriptsize$0.6$} \psfrag{0.7}{}
\psfrag{0.8}{\hspace{-1mm}\scriptsize$0.8$} \psfrag{0.9}{}
\psfrag{1}{\hspace{-1mm}\scriptsize$1$}
\psfrag{32}{\hspace{-1.5mm} \scriptsize$32$} \psfrag{64}{}
\psfrag{96}{\hspace{-1.5mm} \scriptsize$96$} \psfrag{128}{}
\psfrag{160}{\hspace{-1.5mm} \scriptsize$160$} \psfrag{192}{}
\psfrag{224}{\hspace{-1.5mm} \scriptsize$224$} \psfrag{256}{}
\psfrag{50}{} \psfrag{75}{\hspace{-1.5mm} \scriptsize$75$}
\psfrag{100}{} \psfrag{125}{\hspace{-1.5mm} \scriptsize$125$}
\psfrag{150}{} \psfrag{175}{\hspace{-1.5mm} \scriptsize$175$}
\psfrag{200}{} \psfrag{225}{\hspace{-1.5mm} \scriptsize$225$}
\psfrag{250}{}
\psfrag{a}{\scriptsize (a)} \psfrag{b}{\hspace{-3mm} \scriptsize (b)}
\psfrag{c}{\hspace{-2.2mm} \scriptsize (c)}\psfrag{d}{\hspace{-3mm} \scriptsize (d)}
\psfrag{e}{\hspace{-1mm} \scriptsize (e)}
\psfrag{m}{\hspace{-1.5mm} \scriptsize$m$} \psfrag{snr}{\hspace{1.5mm}\scriptsize$\varepsilon$}
\psfrag{j}{\hspace{-1.5mm} \scriptsize $i$} \psfrag{pdf}{}
\includegraphics[width=4.1cm,keepaspectratio]{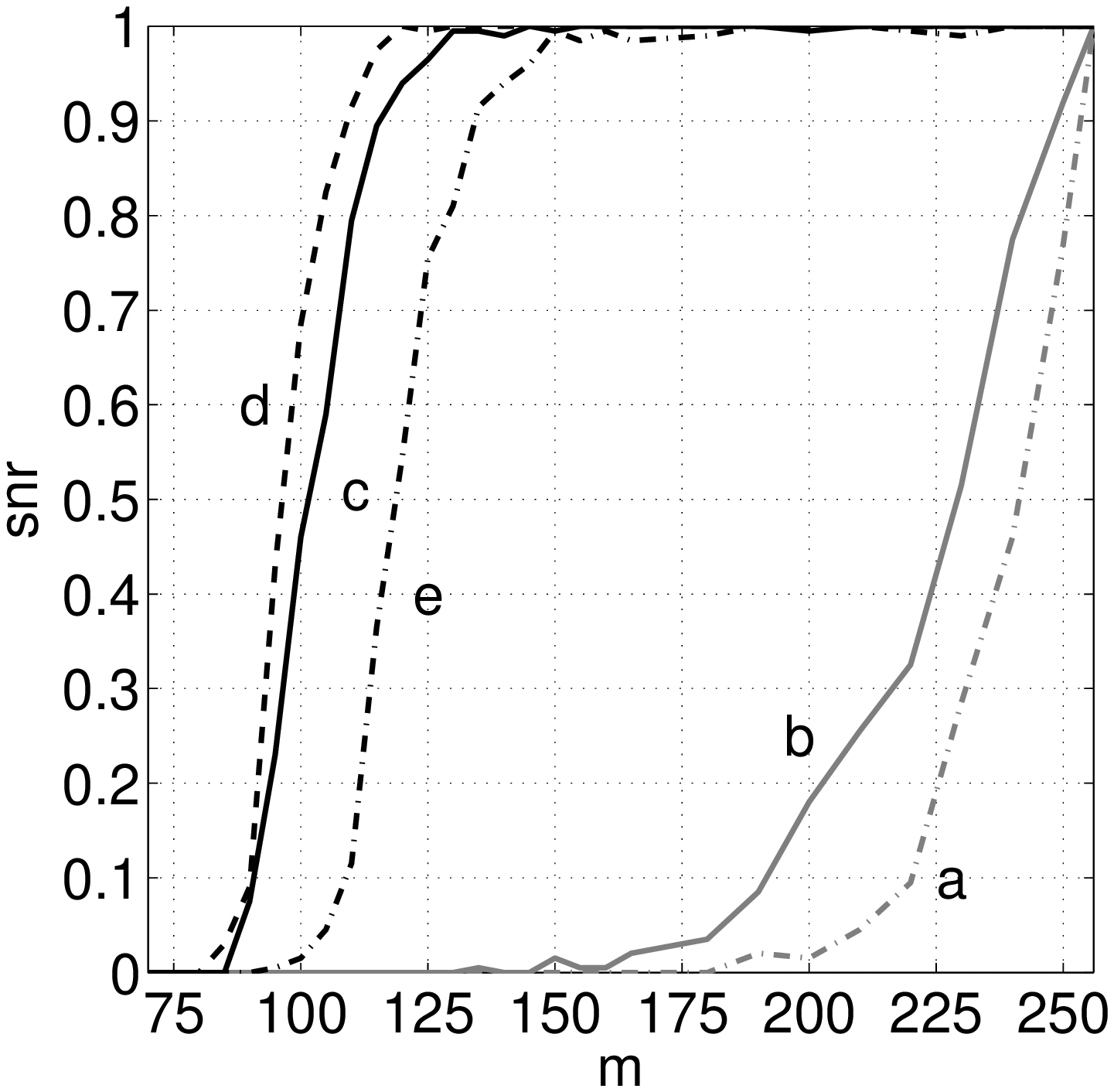}
\includegraphics[width=4.1cm,keepaspectratio]{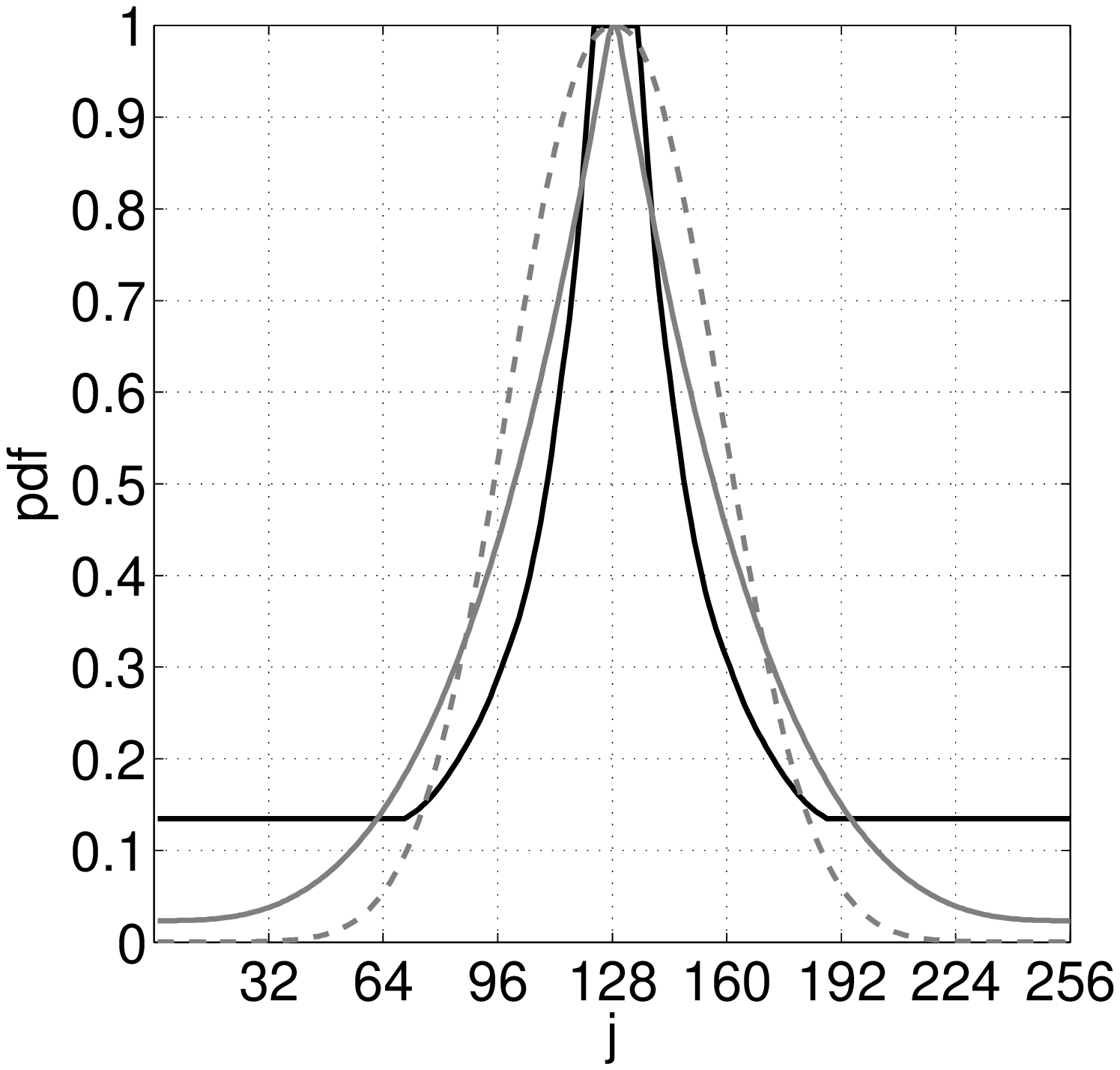}
\caption{\label{fig:MRI}Left panel: probability of recovery $\epsilon$ of a MRI signal as a function of the number of measurements $m$. The light dot-dashed, light continuous, dark dot-dashed, dark continuous and dark dashed curves show, respectively, the probability of recovery obtained with: a uniform profile (a); the optimized profile obtained with the matrix $\ma{B}$ (b); the sampling profile proposed in \cite{wang10} (e); the optimized profile obtained with the matrix $\ma{C}$ (c); the typical MRI profile used in \cite{lustig07} (d). Right panel: Optimized sampling profile obtained with the matrix $\ma{C}$ for $m=70$ (dark continuous curve) in comparison with the one used in MRI (light continuous curve) and the one proposed in \cite{wang10} (light dashed curve).\vspace{-2mm}}
\end{figure} 

In order to evaluate our method when prior information is available on the support $S$, we perform a simplified MRI experiment. In this perspective, an \emph{in vivo} brain image of size $256 \times 256$ was acquired on a $7$ Tesla scanner (Siemens, Erlangen, Germany). As suggested in \cite{lustig07}, we consider a Daubechies-$4$ wavelet basis as sparsity basis and decompose each line\footnote{Lines without any signal (background) are withdrawn. After this operation, $151$ lines are left.} of the brain image into this basis. The resulting vectors are then hard-thresholded at $s=50$. All vectors but one are seen as a data set providing prior information on the support $S$ of typical MRI signals. The average of the values $\sum_{j \in S} \left| \scp{\bm \phi_i}{\bm \psi_j} \right|^2$, for each $i$ in $\{1, \ldots, N\}$, serves to create the matrix $\ma{C}$. The remaining signal, not considered in the data set, is considered as the signal under scrutiny, probed according to relation (\ref{eq:measurement model}) and reconstructed from different number of measurements $m$ by solving the $\ell_1$-minimization problem. For each value of $m$, the selected Fourier basis vectors are chosen using the method described in Section \ref{sec:variable density sampling} with: a uniform density profile (a); the optimized sampling profile $\hat{\bm{p}}$ obtained with $\lambda = 0.05$ and the matrix $\ma{B}$ (b) or with the matrix $\ma{C}$ (c); a typical sampling profile used in MRI (d) \cite{lustig07}; the sampling profile\footnote{Note that the intrinsic parameters the sampling profiles (d) and (e) are manually chosen to obtained the best reconstructions.} proposed in \cite{wang10} (e).

Figure \ref{fig:probability recovery} shows the probability of recovery $\varepsilon$ of $s$-sparse signals as a function of the number of measurements for the first two experiments. The probability of recovery obtained with the spread spectrum technique is also presented \cite{puy11a, puy11b}. Note that this technique, as for random Gaussian matrices, was proved to be universal, i.e., the number of measurements for the recovery of sparse signals is reduced to its minimum independently of the sparsity basis. One can note that the probability of recovery with the optimized sampling is always better than with the uniform sampling. With a sampling in the Fourier basis, one can also note that the recovery becomes almost optimal. Indeed, the number of measurements needed to reach a probability $1$ of recovery is almost the same with the spread spectrum technique and with an optimized profile. These results confirm our theoretical predictions and illustrate the efficiency of variable density sampling.

As illustration, Figure \ref{fig:probability recovery} also shows optimized sampling profiles obtained for the two sensing bases and $m=300$ as well as the corresponding values of the diagonal entries of the matrix $\ma{B}$. One can note that the shapes of the sampling profiles are highly correlated to the values in the matrix $\ma{B}$.

Figure \ref{fig:MRI} shows the probability of recovery $\varepsilon$ of the MRI signal as a function of the number of measurements. One can note that with the uniform sampling (a), the signal is recovered with probability $1$ only when $m=N$. The results are slightly improved with the optimized profile (b) obtained with the matrix $\ma{B}$. The sampling profiles (c), (d), and (e) drastically enhance the performance. Our optimized profile (c) obtained with the matrix $\ma{C}$ performs better than the profile (e) and similarly to the profile (d) typically used in MRI. These results provide a theoretical underpinning to VDS procedures used in MRI. It also shows that the refinement proposed for our technique can drastically enhances the performance of compressed sensing in practical applications.

For illustration, Figure \ref{fig:MRI} also shows the sampling profiles (c), (d), and (e) for $m=70$. One can notice that the profiles (c) and (d) are very similar to each other. This explains again the effectiveness of VDS profiles commonly used in MRI.

%
%
%
%
\section{Conclusion}
\label{sec:conclusion}

In the aim of optimizing variable density sampling profiles in the context of compressed sensing, we have introduced a minimization problem for the coherence between the sparsity and sensing bases. This problem is solved with the use of convex optimization algorithms. We have also discussed a refinement of our technique when prior information is available on the signal support in the sparsity basis. The effectiveness of the method is confirmed by numerical experiments. In particular, for signals sparse in a wavelet basis and probed in the Fourier domain, simulations show that our technique leads to optimal recovery. Indeed our technique gives similar probabilities of recovery as the spread spectrum method recently proved to be optimal. Our results also provide a theoretical underpinning to VDS procedures used in MRI.

%
%
%
%
\appendices

\section{}
\label{ap:proof coherence}

The proof of this theorem follows exactly the method used to prove Theorem $4.2.$ in \cite{rauhut10}. The only difference resides in the estimate of the singular values of the operator $\ma{A}^\dagger_{\Omega S}\ma{A}_{\Omega S}$ (see Theorem $7.3$, \cite{rauhut10}), where $\ma{A}_{\Omega S} \in \Cbb^{m \times s}$ is the restriction of the matrix $\ma{A}_{\Omega}$ to the columns indexed by $S$.
\begin{lemma}
Let $A = \ma{\Phi^\dagger\Psi} \in \Cbb^{N \times N}$, $\bm{p} = \left\{p_i\right\}_{1 \leq j \leq N} \in \set{P}(m)$, $\delta \in (0, 1/2]$, and define $\ma{P} \in \Rbb^{N \times N}$ the diagonal matrix with entries $p_i^{1/2}$ on the diagonal, $1 \leq i \leq N$. Assume that the $m$ measurement vectors are selected according to $\bm{p}$ and suppose that $s \geq 2$. For a universal constant $C>0$, the normalized matrix $\tilde{\ma{A}} = \ma{P^{-1}A}$ satisfies $\norm{\tilde{\ma{A}}^\dagger_{\Omega S}\tilde{\ma{A}}_{\Omega S} - \ma{I}} \leq \delta$ with probability at least $1 - 2^{3/4} s \exp\left[-\frac{m \delta^2}{C N \, \mu^2(\bm{p}) s} \right].$
\end{lemma}
\begin{IEEEproof}
Let us denote $\ma{Y} = \tilde{\ma{A}}^\dagger_{\Omega S}\tilde{\ma{A}}_{\Omega S} - \ma{I} = \sum_{i=1}^{N} \delta_i \tilde{\bm{a}}_i^\dagger \tilde{\bm{a}}_i - \ma{I} \in \Cbb^{s \times s}$ where $\tilde{\bm{a}}_i  \in \Cbb^{1 \times s}$ is the $i^{\rm th}$ row of $\tilde{\ma{A}}_{S}$. The proof starts by noticing that $\E\left[Y\right] = \sum_{i=1}^{N} p_i \tilde{\bm{a}}_i^\dagger \tilde{\bm{a}}_i - \ma{I} = \sum_{i=1}^{N} \bm{a}_i^\dagger \bm{a}_i - \ma{I} = \ma{A}^\dagger_{S}\ma{A}_{S} - \ma{I} = 0$. We can thus continue with the use of a symmetrization technique to bound the expected value of the norm of $\ma{Y}$ (see Lemma $6.7$ in \cite{rauhut10}, or proof of Theorem $3.1$ in \cite{candes07}). Let $(\epsilon_1, \dots, \epsilon_N)$ be a Rademacher sequence independent of $\left(\delta_1, ..., \delta_N\right)$ and $p \geq 2$, then $\E \norm{\ma{Y}}^p \leq  2^p \, \E \norm{\sum_{1 \leq i \leq N} \epsilon_i \delta_i \tilde{\bm{a}}_i^\dagger \tilde{\bm{a}}_i}^p$. Noticing that $\tilde{\ma{A}}_{\Omega S}$ has at most rank $s$, using Fubini's theorem, Rudelson's lemma (see Lemma $6.18$, \cite{rauhut10}) conditional on $\left(\delta_1, \dots, \delta_N \right)$, and the Cauchy Schwarz inequality yields
\begin{eqnarray*}
\E \norm{\ma{Y}}^p \leq 2^{3/4 + p} \, s \, \left(\frac{p}{{\rm e}}\right)^{p/2} \sqrt{\E \norm{\tilde{\ma{A}}_{\Omega S}^\dagger \tilde{\ma{A}}_{\Omega S}}^p \, \E \left[ \max_{i : \delta_i = 1} \norm{\tilde{\bm{a}}_i}_2^{2p} \right]}.
\end{eqnarray*}
The previous equation is identical to equation $(7.6)$ in the proof of Theorem $7.3$ in \cite{rauhut10}. We can follow the same remaining steps of this proof to terminate ours. We still need however to provide a bound on $\max_{i : \delta_i = 1} \norm{\tilde{\bm{a}}_i}_2^2$. If the support $S$ is fixed and known in advance, we have
\begin{eqnarray}
\label{eq:bound model}
\max_{i : \delta_i = 1} \norm{\tilde{\bm{a}}_i}_2^2 \leq  \max_{1\leq i \leq N} \frac{\sum_{j \in S} \left| \scp{\bm \phi_i}{\bm \psi_j} \right|^2}{p_i} = \frac{sN}{m} \mu(\bm{p}, S)^2.
\end{eqnarray}
In the general case where $S$ is unknown, we can write
\begin{eqnarray}
\label{eq:bound no model}
\max_{i : \delta_i = 1} \norm{\tilde{\bm{a}}_i}_2^2 \leq s \max_{1\leq i,j \leq N} \frac{\left| \scp{\bm \phi_i}{\bm \psi_j} \right|^2}{p_i} = \frac{sN}{m} \mu(\bm{p})^2.
\end{eqnarray}
\end{IEEEproof}

\section{}
\label{ap:proof prox}

The proximity operator of $\gamma \norm{\ma{B}\,\cdot}_\infty$, $\gamma > 0$, is the unique solution of ${\rm prox}_{\gamma\norm{\ma{B}\,\cdot}_\infty}(\bm{q}) = \argmin_{\bm{x} \in \Rbb^N} 1/2 \norm{\bm{q} - \bm{x}}_2^2 + \gamma \norm{\ma{B}\,\bm{x}}_\infty$.
\begin{proposition}
For any $\bm{q} \in \Rbb^N$, with $B \in \Rbb^{N \times N}$ defined as above, we have ${\rm prox}_{\gamma \norm{\ma{B}\,\cdot}_\infty}(\bm{q}) = \bm{q} - \gamma\,{\rm proj}_{\set{C}}(\bm{q}/\gamma)$, where $\set{C} = \left\{x \in \Rbb^N :  \norm{\ma{B}^{-1}x}_1\leq1\right\}$ and ${\rm proj}_{\set{C}}$ denotes the projection onto the set $\set{C}$.
\end{proposition}
\begin{IEEEproof}
From Theorem $14.3$ in \cite{bauschke11}, we have $\bm{q} = {\rm prox}_{\gamma \norm{\ma{B}\,\cdot}_\infty}\left(\bm{q}\right) + \gamma \, {\rm prox}_{\gamma^{-1}\norm{\ma{B}\,\cdot}_\infty^*}\left(\bm{q}/\gamma\right)$, for all $\bm{q} \in \Rbb^N$. In the previous relation, $\norm{\ma{B}\,\cdot}_\infty^*$ denotes the Fenchel conjugate of $\norm{\ma{B}\,\cdot}_\infty$. As $\ma{B}$ is a bijection (it is a diagonal matrix with strictly positive entries), one can show that $(\norm{\ma{B}\,\cdot}_\infty)^* = \iota_{\set{C}}(\cdot)$ where $\iota_{\set{C}}$ denotes the indicator function of the set $\set{C}$ (Proposition $13.20$, \cite{bauschke11}). Finally, we have ${\rm prox}_{\gamma^{-1}\norm{\ma{B}\,\cdot}_\infty^*} = {\rm prox}_{\gamma^{-1}\iota_{\set{C}}(\cdot)} = {\rm proj}_{\set{C}}$. Combining the last result with the first relation terminates the proof.
\end{IEEEproof}
%

%
%
%
%

 \end{document}